# SUPER YANG-MILLS THEORY AS A RANDOM MATRIX MODEL

W. Siegel[1]

*Institute for Theoretical Physics*
*State University of New York, Stony Brook, NY 11794-3840*

**ABSTRACT**

We generalize the Gervais-Neveu gauge to four-dimensional N=1 superspace. The model describes an N=2 super Yang-Mills theory. All chiral superfields (N=2 matter and ghost multiplets) exactly cancel to all loops. The remaining hermitian scalar superfield (matrix) has a renormalizable massive propagator and simplified vertices. These properties are associated with N=1 supergraphs describing a superstring theory on a random lattice worldsheet. We also consider all possible finite matrix models, and find they have a universal large-color limit. These could describe gravitational strings if the matrix-model coupling is fixed to unity, for exact electric-magnetic self-duality.

---

[1]　Internet address: siegel@insti.physics.sunysb.edu.

# 1. HIGGS MODELS FOR STRING THEORIES

String theory originated from dual models. This type of duality occurs in any relativistic theory that has (complete) Regge behavior (analyticity in angular-momentum space) at the tree level [1]. "Trees" are defined as graphs that have only poles in momentum space, and not cuts. They can therefore be used to define a local lagrangian, from which loops follow in the usual way. Such a string field theory lagrangian has an infinite number of (particle) fields.

Regge behavior is physically the property that all physical states act as bound states [2]. We can therefore consider, instead of a string field theory with an infinite number of fields, a particle field theory with a finite number of fields, from which the physical states arise as bound states. In particular, a hadronic string theory is expected to result from confinement in quantum chromodynamics (QCD).

When a fundamental particle in field theory can be identified as one of the bound states by appearing on a Regge "trajectory", it is said to have "Reggeized". The only theories for which all fundamental particles in the theory are known to Reggeize are Higgs models where all vectors are nonabelian, all particles are massive, and certain restrictions are imposed on the representations of the scalars (which are satisifed in particular if the theory is supersymmetric) [3]. (Although Regge trajectories occur more generally, it often happens that only some states act as bound, while others do not have Regge behavior, so that Reggeization of the theory is not complete. Such theories are not suitable for describing strings. Also, Reggeization of massless states is not understood because of infrared divergences.)

In any nonabelian theory the physical states are described by composite gauge-singlet fields, and for Higgs models the fundamental fields can be replaced by these composite fields through local field redefinitions equivalent to gauge transformation to the "unitary gauge". The gauge-covariant way to describe the Higgs mechanism is then not to say that the "gluon" has become massive, but rather that a massive vector has arisen as a color-singlet bound state of a gluon and scalar "quarks". This identification of a tree-level state in terms of a composite gauge-singlet field gives a natural explanation of Reggeization, since the other bound states on the Regge trajectory can be identified with excitations of this same composite field.

Confinement and the Higgs mechanism can be related by electric-magnetic duality transformations, so a QCD-like theory with confinement can be reformulated as a Higgs theory whose scalars were magnetic monopoles in the original theory [4]. Such duality transformations also replace the coupling constant with its inverse, so weak coupling in one formulation is strong coupling in the other, and the perturbation expansions are different. However, the coupling constant of the corresponding string theory is not simply related, being more like the inverse of the number of colors, so neither formulation seems to have preference in its relationship to strings. (Replacing the coupling constant of the field theory with its inverse is actually related to changing the sign of the cosmological term of the string.) Thus, it is possible to consider nonabelian Higgs models as alternatives to QCD, at least for the purpose of studying general features of string theories (including hadronic strings).

One way used to study bound states is to sum an infinite subset of graphs. In particular, "ladder" graphs have been used in various gauge and nongauge theories to study Regge behavior and Reggeization [2]. Another example is summing over "fishnet" graphs in massive scalar field theory as an approximation for the derivation of the usual bosonic string as bound states, with the fishnets representing a square lattice for the world sheet of the string [5]. In this approach, the use of strings to describe the "theory of everything" cannot be called a "theory of fundamental strings", since these strings are treated as composite in the same way as hadronic strings. (This could lead to considering hadronic strings as bound states of quark-gluon strings, which in turn are bound states of preonic strings, which are themselves bound states of...)

A more complete method is to sum all graphs, but treat a certain subset as the lowest order in a systematic perturbation expansion. The large-n expansion for group U(n) is such an approach that automatically produces a string-like perturbation expansion by identifying the topology of the Feynman graphs of the field theory with those of string theory [6]. (Similar methods may often be used for SO(n) and USp(2n).) This method of deriving strings as bound states was first explicitly applied to two-dimensional QCD [7]. More recently it has been applied to massive scalar field theory, as a refinement of the fishnet approach, to show that the usual bosonic string arises in an appropriate limit. The geometry of the scalar Feynman graphs (which are identified with surfaces by the large-n expansion) is equated with that of the world sheet on a random lattice, so summing over graphs is the same as summing over world-sheet metrics [8]. (The fishnet graphs are the subset of graphs corresponding to the conformal gauge for the metric, but ignoring ghosts.)

In earlier papers we considered the generalization of this random lattice method to superstrings [9,10]. The approach taken was to start with first-quantization of the superstring on a random lattice and look for a second-quantized field theory that generated the same amplitudes. In this paper we will apply the reverse approach. Since the superstring has critical dimension ten, the natural candidate for such a theory is super Yang-Mills theory: Supersymmetry in ten dimensions requires maximum spin one or higher, and Yang-Mills theory, unlike supergravity, can naturally be associated with the group U(n) as the gauge group for arbitrarily large n. (Of course, this theory is not purely nonabelian because of the U(1) factor, but we will assume it is close enough for large n.) Super Yang-Mills theory is also a better candidate for obtaining a consistent string theory: In four dimensions (or less), massless Yang-Mills theory (probably) has confinement and massive Yang-Mills theory has Reggeization, so we can expect to get a theory with only bound states, while the same is not true for scalar theories, used to derive the bosonic string.

Since superspace is not understood in ten dimensions, we will consider only four dimensions. (Even six-dimensional superspace is not understood nearly as well, although we will make brief remarks on it.) This technical difficulty can be overcome by using the four-dimensional superspace formalism for ten-dimensional supersymmetry, but we will ignore these complications for simplicity. Of course, the four-dimensional theory is more interesting for physical reasons. Also, confinement and Reggeization [3] are closely related to asymptotic freedom, and thus seem unlikely in higher dimensions. (In fact, the scalar random matrix model for the bosonic string, if we choose four dimensions and the vertex to be four-point, describes wrong-sign $\phi^4$ theory, which is also asymptotically free.) We can also consider superstring theories in four dimensions, with extra degrees of freedom (perhaps derived from fermionization of six $x$'s) to cancel the conformal anomaly.

To allow identification of a field theory as a random matrix model for string theory, it must have a propagator that is simply $1/(p^2 + M^2)$. This propagator is then approximated as $M^{-2} e^{-p^2/M^2}$. Thus, mass is as important for random matrix models as it is for other methods used in studying Reggeization. This exponential factor arises directly from the usual $(\partial x)^2$ term in the string action $S$ of the first-quantized functional integral of $e^S$. For the present case this implies that we must use only the real scalar superfield $V$ normally used to describe super Yang-Mills, but not the chiral scalar superfields $\phi$ used to describe ghost and matter multiplets. Although ghosts can be avoided in unitary gauges, such gauges have "nonrenormalizable" propagators, rather than the "renormalizable" propagator $1/(p^2 + M^2)$. However, Gervais and Neveu long ago gave a gauge with renormalizable propagators in certain (bosonic) Higgs models, for which the scalar and ghost contributions exactly cancel to all loops [11]. The resulting theory is described completely by a vector field, with the Higgs scalar state appearing as the fourth, timelike polarization of this massive vector. Perhaps not coincidentally, they discovered this gauge from string theory.

We will supersymmetrize this gauge and model. The main features of this theory are: (1) the propagator is the usual renormalizable massive propagator $1/(p^2 + M^2)$, (2) the theory is described completely in terms of an unconstrained hermitian matrix (super)field that is a scalar, (3) the three-point vertex includes a nonderivative term, and (4) the theory is N=2 supersymmetric. Although the bosonic Gervais-Neveu model shares the first feature, the first three features are necessary to obtain a random matrix model of a string theory.

We also discuss the possibility of string theories with matrix models that are finite (conformally invariant, perhaps up to mass terms). We consider all conformal four-dimensional theories (N=2 theories with the right matter to cancel the $\beta$ function, including the N=4 case). We find they all have the same large-n limit. Furthermore, it seems such a theory is required in order to describe strings with massless states (such as the graviton), and the existence of such states fixes the value of the matrix-model Yang-Mills coupling constant to be unity, corresponding to exact electric-magnetic self-duality. (On the other hand, our explicit supersymmetric Gervais-Neveu model has a nonvanishing $\beta$ function, and so apparently describes a hadron-like string.)

## 2. COMPLEX GAUGES

Gervais and Neveu derived the Feynman rules for the massless fields resulting from the low-energy limit of bosonic string theory, and the corresponding rules for these fields when their mass has been shifted from zero. (This shift is inconsistent with unitarity for the string theory, but consistent for the field theory obtained by taking the low energy limit for the trees before unitarizing.) They then explained how these unusual rules, which are simpler than those usually given for these theories, can be derived by field theory methods, without reference to string theory. Here we will not repeat the string theory part of the analysis, but just give a simplified version of this gauge fixing.

We first consider the case of pure massless Yang-Mills theory. (Recently this case has been used in describing simplified rules for QCD [12].) We start with the gauge-invariant lagrangian:

$$L_0 = -\tfrac{1}{4}F^2, \quad F_{ab} = -i[\nabla_a, \nabla_b], \quad \nabla_a = \partial_a + iA_a$$

where $A$ is an n×n matrix describing gauge group U(n), and we use the convention that the action $S = \frac{1}{g^2} \int d^4x \, tr \, L$ appear as $e^{+S}$ in the functional integral (so this $S$ is nonpositive definite). We then choose the complex gauge-fixing function

$$\Psi_0 = \partial \cdot A + iA^2$$

(Similar results are obtained for $\Psi_0 = \partial \cdot A - iA^2$.) By any of the usual gauge-fixing procedures for Fermi-Feynman gauges, this results (after some algebra) in the gauge-fixed lagrangian

$$L_A = -\tfrac{1}{4}F^2 - \tfrac{1}{2}\Psi_0^2 = \tfrac{1}{2}A \cdot \Box A + 2iA^a A^b \partial_b A_a + \tfrac{1}{2}A^a A^b A_a A_b$$

and the terms for the ghost $C$ and antighost $\tilde{C}$

$$L_C = \tilde{C}\nabla^2 C - i\tilde{C}C\Psi_0$$

where we have collected terms into covariant ones involving only the covariant derivative $\nabla$, and the remainder.

To generalize to the massive case (nonabelian Higgs model), we couple to scalars that are in the fundamental representation of the U(n) gauge group, as well as the fundamental representation of a global U(n), resulting in a complex n×n matrix with $2n^2$ real components, the same as the number of components of the ghosts plus antighosts. Fixing the quartic self-coupling so the masses of the scalar and vector will come out the same, the new terms in the gauge-invariant lagrangian are

$$L_\phi = \phi^\dagger \nabla^2 \phi - \tfrac{1}{2}R^2, \quad R = \phi^\dagger \phi - \tfrac{1}{2}M^2$$

If we were to use the unitary gauge, we would expand about the vacuum as $\phi = (\phi' + M)/\sqrt{2}$ in the gauge $\phi' = \phi'^\dagger$. (The vacuum value of $\phi^\dagger \phi$ is proportional to the identity.) This is the same as separating $\phi$ into its unitary and hermitian parts as $\phi = U(\phi' + M)/\sqrt{2}$ (where $\phi' \equiv \sqrt{2\phi^\dagger \phi} - M$) and using a gauge transformation with this $U^{-1}$ as a field redefinition.

For a Gervais-Neveu-type gauge we instead modify the previous gauge-fixing function to

$$\Psi = \Psi_0 + iR$$

This choice has the interesting feature of cancelling the scalar self-interaction completely (including the mass term), while leaving the ghost terms unmodified from the pure Yang-Mills case. ($R$ is gauge invariant.) The final result for the total lagrangian is

$$L = L_0 + L_\phi - \tfrac{1}{2}\Psi^2 + L_C$$
$$= (L_A - \tfrac{1}{2}M^2 A^2) + (\phi^\dagger \nabla^2 \phi - i\phi^\dagger \phi \Psi_0) + (\tilde{C} \nabla^2 C - i\tilde{C} C \Psi_0)$$

Since the scalar lagrangian is identical in form to that of the ghosts, they exactly cancel in loops to all orders. Note that both the scalars and the ghosts in this lagrangian are massless, while the vector is massive. This is a reflection of the fact that both the scalar and ghost fields describe unphysical polarizations, while the vector field describes all the physical ones. This effect is expected from the linearized form of the gauge condition:

$$\Psi \approx \partial \cdot A + iM\phi'$$

so $\Psi = 0 \Rightarrow p \cdot A \approx M\phi'$. Then the Feynman rules for this theory of vectors plus scalars can be described completely by the lagrangian $L_A - \tfrac{1}{2}M^2 A^2$ in terms of just a vector field.

Gervais and Neveu used a slightly different analysis that looked more like the "$R_\xi$" renormalizable gauges that have been used in nonabelian Higgs models. In such gauges, one first shifts the Higgs fields by their classical vacuum values, and then picks a gauge-fixing term to cancel the scalar-vector mass crossterms. In practice such gauges are a waste of time in loop calculations, since (1) the same result can be obtained without the shifting and corresponding gauge-fixing modification, since the quantum fields are dummy variables, and (2) the shifting and thus the choice of gauge has to be done all over again due to quantum corrections to the Higgs vacuum values. The easiest way to do gauge fixing in ordinary Higgs models is to use the background field method and do all shifting after calculating the effective action. This means there is no shifting involved in the renormalizable gauge for the quantum fields, while all shifting is done on the background fields, for which one can even use the unitary gauge. However, for the present model the modified Gervais-Neveu gauge allows an even simpler treatment, since the number of fields is reduced.

The way that string-inspired complex gauges simplify the massive Yang-Mills propagator is analogous to the way string theory simplifies the graviton propagator. Both cases can be described as a conformal theory (in four dimensions) coupled to two scalars. In the Yang-Mills case, the two scalars are the one eaten by the vector, and the physical Higgs scalar; in the gravity case, they are the one eaten by the conformal graviton to produce the Einstein graviton, and the physical scalar usually called the "dilaton". (In reality, the conformal compensating scalar eaten by the graviton is the true dilaton, since it couples universally to the trace of the energy-momentum tensor.)

## 3. SUPERSYMMETRY AND RANDOM MATRICES

The generalization to the supersymmetric case involves replacing the vector with a vector multiplet (vector + Weyl spinor), and the scalars (including ghosts) with scalar multiplets (complex scalar + Weyl spinor). For the massive case we want to end up with just a real scalar superfield, which should contain only physical polarizations. This means 8+8 fermionic+bosonic polarizations, while the massless vector and scalar multiplets both contain 2+2. Since the real scalar superfield has room for only one vector, the counting for the whole superfield is equivalent to that for one vector and three scalar multiplets. This result is familiar from background field quantization of super Yang-Mills [13], where the ghosts consist of three scalar multiplets, which are necessary to cancel the three unphysical scalar multiplet degrees of freedom in the real scalar superfield. This result also follows from noting that first quantization of the superparticle tells us that only one quarter of the fermionic coordinates are physical [14] (half are killed by first-class constraints, and half of the remainder by second-class ones), and thus a superfield with arbitrary dependence on all coordinates has the same component count as an N=4 superfield, which in this case means N=4 super Yang-Mills theory. Similarly, the light-cone superfield for N=4 super Yang-Mills theory [15] is a single superfield that is a function of four anticommuting coordinates. N=4 super Yang-Mills theory is also described by N=1 coupled to three scalar multiplets [16].

However, from the bosonic case we know that two of the three scalar multiplets must describe the fundamental representation of the gauge group, and of the global group: Two multiplets because in the bosonic case we started with twice as many scalars as vectors, since the scalars were a complex representation while the vectors were a real one. The remaining one scalar multiplet must therefore be a real representation, namely the adjoint. This N=1 multiplet structure is the same as that for N=2 super Yang-Mills theory coupled to a single N=2 scalar multiplet [16] in the fundamental×fundamental representation of the two U(n)'s. We therefore consider this N=2 supersymmetric theory, with the mass scale introduced by an N=2 Fayet-Iliopoulos term:

$$L_0 = \int d^2\theta \ W^2 + \int d^4\theta \ (e^{-V}\bar{\phi}_0 e^V \phi_0 + \bar{\phi}_+ e^V \phi_+ + \phi_- e^{-V}\bar{\phi}_-)$$

$$+ \left[\int d^2\theta \; \phi_0(\phi_+\phi_- - \tfrac{1}{2}M^2) + h.c.\right]$$

where $W^2 = \tfrac{1}{2}W^\alpha W_\alpha$, $W_\alpha = i\bar{d}^2 e^{-V} d_\alpha e^V$, $V$ is hermitian, and the three $\phi$'s are chiral. The linear $\phi_0$ term, plus its complex conjugate, is equivalent to the usual N=1 Fayet-Iliopoulos term $\int d^4\theta \; V$ by N=2 supersymmetry: These three terms are the triplet of auxiliary fields for the N=2 vector multiplet under the internal SU(2) of the N=2 supersymmetry. We find our particular choice from these three terms the most convenient for gauge fixing. This action also directly corresponds to a six-dimensional N=1 supersymmetric one.

The gauge-fixing terms for the supersymmetrized Gervais-Neveu-type gauge are

$$L_1 = \int d^2\theta \; \chi(\bar{d}^2 e^{-V} + \phi_0) + \int d^2\bar{\theta} \; \bar{\chi}(d^2 e^V + \bar{\phi}_0)$$

where $\chi$ is a chiral Lagrange multiplier. The two terms are not complex conjugates, just as the bosonic $L_1$ was not real. They break invariance under charge conjugation $V \to -V^T$, $\phi_0 \to \phi_0^T$, $\phi_\pm \to \phi_\mp^T$. However, they preserve parity invariance $V \to -V$, $\phi_0 \leftrightarrow \bar{\phi}_0$, $\phi_\pm \leftrightarrow \bar{\phi}_\mp$, $d \leftrightarrow \bar{d}$. (The same is true for both invariances in the bosonic case.) The linear $V$ terms in the expansion of the exponentials contain the $\partial \cdot A$ part of the bosonic Gervais-Neveu gauges, and the $V^2$ terms contain the $A^2$ terms. The corresponding ghost terms can be written as

$$L_C = \int d^4\theta \; (\overline{\tilde{C}} e^V C + \tilde{C} e^{-V} \bar{C}) + \left(\int d^2\theta \; \tilde{C}\phi_0 C + h.c.\right)$$

where we have used the equation of motion for $\phi_0$ implied by the Lagrange multiplier. These terms are identical to the terms for $\phi_\pm$ under the identification

$$(\phi_+, \phi_-, \bar{\phi}_+, \bar{\phi}_-) \quad \leftrightarrow \quad (C, \tilde{C}, \overline{\tilde{C}}, \bar{C})$$

Thus, we again can drop all terms involving the N=2 scalar ghost multiplet $(C, \tilde{C})$ as well as the N=2 scalar matter multiplet $\phi_\pm$, while keeping all terms involving the N=2 vector multiplet $(V, \phi_0)$, but $\phi_0$ is determined in terms of $V$ after eliminating the Lagrange multiplier $\chi$ by its equation of motion. The net result is then

$$L = \int d^2\theta \; W^2 + \int d^4\theta \; [e^{-V}(d^2 e^V)e^V \bar{d}^2 e^{-V} + \tfrac{1}{2}M^2(e^V + e^{-V})]$$

We can further simplify the action by the field redefinition

$$e^V \to 1 + V$$

(This is legal at least for purposes of perturbation theory.) This also simplifies the gauge tranformation law of $V$: Previously it was nonpolynomial for $V$, but simply $\delta e^V = i\bar\Lambda e^V - e^V i\Lambda$ in terms of $e^V$. Now it is just

$$\delta V = i(\bar\Lambda - \Lambda) + i(\bar\Lambda V - V\Lambda)$$

which is very similar to the bosonic transformation, having just abelian and homogeneous terms. The lagrangian now becomes

$$L = \int d^4\theta \ \left[ -\tfrac{1}{2}\frac{1}{1+V}(d^\alpha V)\bar d^2 \frac{1}{1+V} d_\alpha V + \frac{1}{1+V}(d^2 V)(1+V)\bar d^2 \frac{1}{1+V} \right.$$

$$\left. + \tfrac{1}{2}M^2\left(V + \frac{1}{1+V}\right) \right]$$

This redefinition has no effect on the kinetic term, but simplifies the three-point vertex by replacing a $\Box$ with an $M^2$. The kinetic term is the expected

$$-\tfrac{1}{2}\int d^4\theta \ V(\Box - M^2)V$$

The three-point vertex

$$\int d^4\theta \ [-\tfrac{1}{2}M^2 V^3 + (\bar d^{\dot\alpha} V)V i \partial_{\alpha\dot\alpha} d^\alpha V]$$

has the important simplification that the number of spinor derivatives has been reduced from four to two. This is the best one can expect, since otherwise (with no $d$'s) the one-loop four-point function would be trivial. Also, two $d$'s are needed to give the bosonic $AA\partial A$ term, since $A$ appears at quadratic order in $\theta$ in $V$. Unfortunately, by the same argument, the four-point function must have a $d^4$ term to give the bosonic $A^4$ term. However, the three-point vertex may be sufficient for the large-n expansion, since for the bosonic string the result was independent of the form of the potential. (Conversely, we know that, when deriving a field theory from the low-energy limit of string amplitudes, the three-point vertex follows straightforwardly, while higher-point vertices follow from just the three-point string vertex by contractions of propagators.)

An important difference of the three-point vertex from the bosonic case is that, although the mass-independent part of the vertex has been simplified again from two terms to one, there is now also a mass-dependent term with no derivatives. Such a term is in fact required for a random matrix model interpretation [9,10]: Upon obtaining second-quantized Feynman rules by latticizing the first-quantized string action, both the vertex operator (which comes from the string action's Wess-Zumino term) and the propagator (which comes from the rest) appear as exponentials, simply because the string action itself does. (An individual second-quantized Feynman diagram corresponds to a particular geometry for the world-sheet lattice, and therefore to the entire exponential of the first-quantized string action, although it also corresponds to a single term in the expansion of the exponential of a different, second-quantized action.) Thus the two terms in the cubic interaction, just as the two parts of the kinetic term, are treated as the first two terms in the expansion of an exponential.

The next step would be to find explicitly the gauge-invariant, continuum world-sheet action that gives this propagator and vertex upon random-lattice quantization. Since the vertex operator has three derivatives, it corresponds to a gauge-fixed Wess-Zumino term of world-sheet conformal dimension three. (Each derivative corresponds to a vector current on the world-sheet.) In an earlier paper [10] we considered a slightly modified form of the Green-Schwarz superstring, and derived a random matrix model without gauge fixing. The result was similar to the present one except that, in addition to the usual $\theta$ coordinate, there was an anticommuting spinor coordinate $\phi$ (and its derivative $\omega$). The dependence on this coordinate should have been determined by gauge fixing of the corresponding invariance in the string action. In addition, the usual $\kappa$ gauge invariance had not been fixed, which would have reduced the components of $\theta$ by a factor of two. The derivative term in the three-point vertex was of the form $[V, \omega V] dV$. By comparison with the present action, we see that agreement might be obtained if (1) we start with a 4D N=2 string action and fix $\kappa$ symmetry by reducing $\theta$ to N=1, and (2) we fix $\omega \sim \slashed{p} d$. (Compare this to the quantum-mechanical relation $[p, d] \sim \omega$.) Because of the asymmetry of our new three-point vertex in $d$ and $\bar{d}$, it is clear that the world-sheet gauge-fixing conditions must also be complex. (An alternative might be to try and derive this model by gauge fixing Berkovits' modification of the Green-Schwarz action, which is specialized to the N=1 supersymmetry of four dimensions [17]. A messier alternative would be to quantize the Green-Schwarz string by gauge fixing the $\kappa$ symmetry with an infinite pyramid of ghosts, which would appear as coordinates of the random matrix field.)

## 4. CONFORMAL STRINGS?

Although the superfield $V$ has the same spin content as N=4 super Yang-Mills theory, it describes an N=2 theory that is not conformally invariant. In particular, the U(1) factor of the U(n) group is not even asymptotically free, but perhaps it becomes unimportant in the large-n limit. It would be interesting if a string theory could be derived from a conformally invariant theory, perhaps up to mass terms that break the invariance only at low energy. (This might correspond to a string mechanics action whose dominant term was also spacetime-conformally invariant [18].) In particular, we note that N=4 super Yang-Mills theory is the same in the large-n limit as *any* conformal theory; i.e., as any conformal combination of N=2 super Yang-Mills with N=2 matter multiplets.

Because of the n-dependence of the dimensionality of U(n) representations for large n, there are only three possibilities of matter-multipet representations to consider: (1) 2n fundamental representations (twice as many as used in the model above), (2) one adjoint representation, which is the same theory as N=4, and (3) one symmetric plus antisymmetric second-rank tensor representation, i.e., the direct prodcut of two fundamental representations. (We can also separate the symmetric and antisymmetric tensors, and in the large-n limit this is effectively the same as taking half the combined multiplet. Then we can double either multiplet, or add approximately n fundamental representations. These cases are effectively the same as taking linear combinations of the other cases, and will not be discussed separately.) In terms of Chan-Paton factors, the group theory factors of their propagators are described graphically respectively by: (1) a solid (color) line and a broken (flavor) line, (2) two solid lines oriented in opposite directions (as indicated by arrows), and (3) two solid lines oriented in the same direction.

We consider integrating out all the matter multiplets first. All the theories are described by N=2 super Yang-Mills fields minimally coupled to N=2 matter multiplets that have no self-interactions, so it is sufficient to compare graphs with one matter loop and an arbitrary number of external N=2 Yang-Mills lines. This results in an action that includes the classical N=2 Yang-Mills action plus the effective term from this one-loop determinant. The three cases of matter multiplets differ only in the group theory. In all cases the relevant graphs, to leading order in 1/n, are of the form

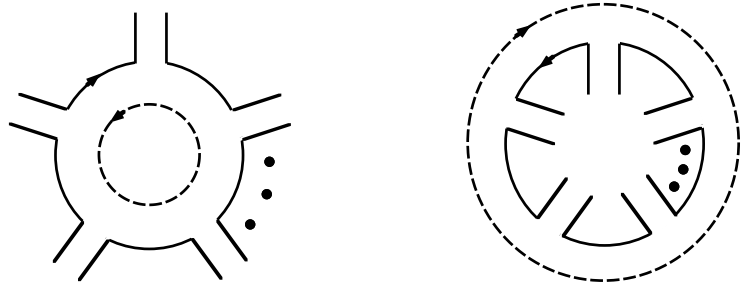

(Graphs lower order in n have N=2 Yang-Mills lines emerging from both the inside and outside of the loop, so there is no continuous circle to give a factor of n.) In the case where the matter multiplet is in the adjoint representation, the broken lines are replaced with solid ones. In the tensor case, the same is true, but the direction of the arrow is reversed. The two diagrams are related by charge conjugation, i.e., reversing the direction of the arrows. We have chosen to orient the diagrams (and thus the world-sheet they define) by choosing all N=2 Yang-Mills lines to go counter-clockwise around the adjacent holes. In the case of the fundamental representation, we have double-counted, since the N=2 gluons couple to only one of the two lines.

In N=1 notation, if we take all external lines to be N=1 Yang-Mills multiplets, then in the left diagram all external factors are $(e^V - 1)$'s, while in the right they are $(e^{-V} - 1)$'s. For the fundamental representation, in the notation used above, the left diagram represents the $\phi_+$ loop, while the right diagram represents the $\phi_-$ loop. Similar remarks apply for the tensor representation, but with a factor of two from coupling to either line. (It has the couplings $e^V \bar{\phi}_+ e^V \phi_+ + \phi_- e^{-V} \bar{\phi}_- e^{-V}$.) For the adjoint representation, both diagrams contribute for the same N=1 matter multiplet, as for the $\phi_0$ part of the N=2 Yang-Mills multiplet, but there are two such N=1 multiplets in the N=2 multiplet, so there is again a factor of two relative to the fundamental representation. The result for the three cases (to leading order in $1/n$) is thus identical (with the number of flavors 2n for the fundamental case). They are also the cases that are finite [19] to leading order in $1/n$, i.e., for the SU(n) subgroup of U(n). We thus have a kind of "universality" for finite theories.

We can also conjecture on the effects conformal invariance and finiteness have on the couplings in the corresponding string theory. Since conformal theories are (probably) self-dual under electric-magnetic duality when replacing the coupling with its inverse [20], there is an exact self-duality when the coupling constant is unity. (The coupling is defined as $ng^2/4\pi$ in the conventions of matrix models and large-n expansion of Yang-Mills fields, which use the normalization of the fundamental representation, versus the $g^2/4\pi$ used in the adjoint representation.) This value of the super Yang-Mills coupling corresponds to vanishing of the world-sheet cosmological constant in the string mechanics action ($\mu_0 = -\frac{1}{2}ln(ng_0^2/4\pi)$). If this exact duality invariance is preserved quantum mechanically, so both the bare and renormalized string cosmological constant vanish, then the string coupling constant is also unrenormalized, and can be identified exactly with the quantized value $1/n$.

Another interesting topic is the appearance of bound-state gravity. Although the graviton, as part of the closed string, naively appears as a bound state in open string field theory, it already appears in the free theory [21], so there is no dynamical mechanism involved. On the other hand, when strings are treated as bound states of random matrix models, the mechanism is nontrivial, requiring an infinite summation of graphs. Thus, random matrix models can be considered as quantum theories of gravity in terms of only a finite number of fields. Of course, such an interpretation requires a string with massless states. In terms of the random matrix model, this just means that the Liouville mode decouples, and so requires that the unrenormalized and renormalized string cosmological constants vanish, which may require a conformal theory. Bound-state gravity is expected to arise naturally from conformal theories [22]. There is evidence from Reggeization arguments to suggest that such a phenomenon may occur in four-dimensional N=4 Yang-Mills theory [23].

## ACKNOWLEDGMENTS

I thank Nathan Berkovits, Jan de Boer, Marc Grisaru, and Martin Roček for discussions. This work was supported in part by the National Science Foundation Grant No. PHY 9309888.

## REFERENCES

[1] G. Veneziano, *Phys. Rep.* **9C** (1974) 199.
[2] T. Regge, *Nuo. Cim.* **14** (1959) 951;
P.D.B. Collins and E.J. Squires, Regge poles in particle physics, Springer tracts in modern physics, v. 45 (Springer-Verlag, Berlin, 1968)
[3] M.T. Grisaru and H.J. Schnitzer, *Phys. Rev.* **D20** (1979) 784, **21** (1980) 1952.
[4] S. Mandelstam, *Phys. Rep.* **23C** (1976) 245, *Phys. Rev.* **D19** (1979) 2391;
G. 't Hooft, *Nucl. Phys.* **B138** (1978) 1.
[5] H.B. Nielsen and P. Olesen, *Phys. Lett.* **32B** (1970) 203;
D.B. Fairlie and H.B. Nielsen, *Nucl. Phys.* **B20** (1970) 637;
B. Sakita and M.A. Virasoro, *Phys. Rev. Lett.* **24** (1970) 1146.
[6] G. 't Hooft, *Nucl. Phys.* **B72** (1974) 461.
[7] G. 't Hooft, *Nucl. Phys.* **B75** (1974) 461.
[8] M.R. Douglas and S.H. Shenker, *Nucl. Phys.* **B335** (1990) 635;
D.J. Gross and A.A. Migdal, *Phys. Rev. Lett.* **64** (1990) 127;
E. Brézin and V.A. Kazakov, *Phys. Lett.* **236B** (1990) 144.
[9] A. Miković and W. Siegel, *Phys. Lett.* **240B** (1990) 363;
W. Siegel, *Phys. Lett.* **252B** (1990) 558.
[10] W. Siegel, *Phys. Rev.* **D50** (1994) 2799.
[11] J.L. Gervais and A. Neveu, *Nucl. Phys.* **B46** (1972) 381.
[12] Z. Bern and D.A. Kosower, *Nucl. Phys.* **B379** (1992) 451;
Z. Bern and D.C. Dunbar, *Nucl. Phys.* **B379** (1992) 562.
[13] M.T. Grisaru, W. Siegel, and M. Roček, *Nucl. Phys.* **B159** (1979) 429;
S.J. Gates, Jr., M.T. Grisaru, M. Roček, and W. Siegel, Superspace, *or* One thousand and one lessons in supersymmetry (Benjamin/Cummings, Reading, 1983) p. 372.

[14] L. Brink and J.H. Schwarz, *Phys. Lett.* **100B** (1981) 310.

[15] S. Mandelstam, *Nucl. Phys.* **B213** (1983) 149;
L. Brink, O. Lindgren, and B.E.W. Nilsson, *Nucl. Phys.* **B212** (1983) 401.

[16] P. Fayet, *Nucl. Phys.* **B149** (1979) 137.

[17] N. Berkovits, *Phys. Lett.* **304B** (1993) 249.

[18] A.M. Polyakov, *Nucl. Phys.* **B268** (1986) 406;
J. Isberg, U. Lindström, B. Sundborg, and G. Theodoridis, *Nucl. Phys.* **B411** (1994) 122;
H. Gustafsson, U. Lindström, P. Saltsidis, B. Sundborg, and R. van Unge, preprint USITP-94-08, hep-th/9410143 .

[19] P.S. Howe, K.S. Stelle, and P.C. West, *Phys. Lett.* **124** (1983) 55.

[20] E. Witten and D. Olive, *Phys. Lett.* **78B** (1978) 97;
L. Girardello, A. Giveon, M. Porrati, and A. Zaffaroni, *Phys. Lett.* **334** (1994) 331, preprint NYU-TH-94-12-1, hep-th/9502057;
N. Seiberg and E. Witten, *Nucl. Phys.* **B431** (1994) 484.

[21] W. Siegel, *Phys. Rev.* **D49** (1994) 4144.

[22] F. Englert, C. Truffin, and R. Gastmans, *Nucl. Phys.* **B117** (1976) 407.

[23] M.T. Grisaru and H.J. Schnitzer, *Nucl. Phys.* **B204** (1982) 267.